\documentclass[a4paper,11pt]{article}
\usepackage{pos}
\usepackage{sidecap}
\usepackage[export]{adjustbox}
\usepackage{capt-of}
\usepackage{caption}

\newcommand*{\cenuns}{CE$\nu$NS}

\title{\cenuns\ at the European Spallation Source}
\ShortTitle{}

\author[a]{A.~Sim\'on}
\onbehalf{on behalf of the $\nu$ESS collaboration}

\affiliation[a]{Donostia International Physics Center (DIPC) and Enrico Fermi Institute, University of Chicago}
\emailAdd{ander.simon@dipc.org}

\abstract{The recent detection of coherent elastic neutrino-nucleus scattering (\cenuns) opens the possibility to use neutrinos to explore physics beyond standard model, with small-size detectors. However, the \cenuns\ process generates signals at the few keV level, requiring very sensitive detector technologies. The European Spallation Source (ESS) has been identified as an optimal source of low energy neutrinos, offering an opportunity for a definitive exploration of all phenomenological applications of \cenuns.

A number of different detector approaches are currently under development for deployment at ESS. These next-generation technologies will be able to observe the process with lower energy threshold and better energy resolution than current detectors. The combination of their observations will allow for a complete phenomenological exploitation of the \cenuns\ signal. In particular, these measurements will not be statistically-limited, a result of the large neutrino flux expected at the ESS.

The main projects currently being developed to detect the \cenuns\ at the ESS are presented: the GanESS project which will use high pressure gas TPC filled with different noble gases; the C$^{\circ}\!$sI project, which employs cryogenic undoped CsI crystals; and p-type point contact HPGe detectors.}

\FullConference{XVIII International Conference on Topics in Astroparticle and Underground Physics (TAUP2023)\\
28.08-01.09.2023\\
University of Vienna}

\begin{document}
\maketitle

\section{Introduction}

Predicted in the 70s, coherent elastic neutrino-nucleus scattering (\cenuns) was only first measured just a few years ago \cite{COHERENT:2017ipa}. In this process low energy neutrinos (few tens of MeV) interact coherently with an atomic nucleus as whole, through the weak current channel, as long as the coherent condition |Q| < 1/R is satisfied, with |Q| being the momentum transfer and R the radius of the nucleus. As a result of the low value of the weak mixing angle for protons, the coupling ends up being effectively proportional to the squared number of neutrons (N$^2$) in the target nucleus. 
The process is of utmost relevance to deepen our knowledge of a large number of neutrino properties. For example, current measurements of this interaction have already been used in nuclear structure studies to improve bounds on Non-Standard neutrino Interactions. Improved \cenuns\ measurements can also be used to better determine the neutrino electromagnetic properties, constrain the weak mixing angle and test both dark matter and sterile neutrinos models. Finally, \cenuns\ can be used to monitor nuclear proliferation thus honing the detection techniques of this process may potentially lead to technological applications.
The most intense neutron spallation source in the coming years, the European Spallation Source (ESS), will produce $\sim$2.53 $\times$ 10$^{23}$ neutrinos per year, almost a factor 10 improvement over the Spallation Neutron Source, where \cenuns\ was first detected. The ESS will provide first protons on target at lower energy in 2025 with the goal to increase it to 5 MW and 2 GeV in the following years. As described in \cite{Baxter:2019mcx,ABELE20231}, the combination of compact detectors and the large neutrino flux produced by the ESS will allow \cenuns\ measurements only limited by few percent systematics and not by statistical fluctuations. Three of the detection techniques  described in \cite{Baxter:2019mcx,ABELE20231} are funded and currently in active development for deployment in the coming years: noble gases time projection chamber, cryogenic pure CsI and p-type high purity Ge detectors. Moreover, the combination of the first two offer a uniquely interesting approach as their response to \cenuns\ is mostly identical given their similarity in nuclear structure, but their technologies and expected systematics are nothing alike. Consequently, the simultaneous use of these detectors will serve to reinforce, or refute, any hint of unexpected observations.

\section{GanESS: Noble gas time projection chamber.}

\begin{figure}[!h]
    \centering
    \includegraphics[width=0.38\textwidth]{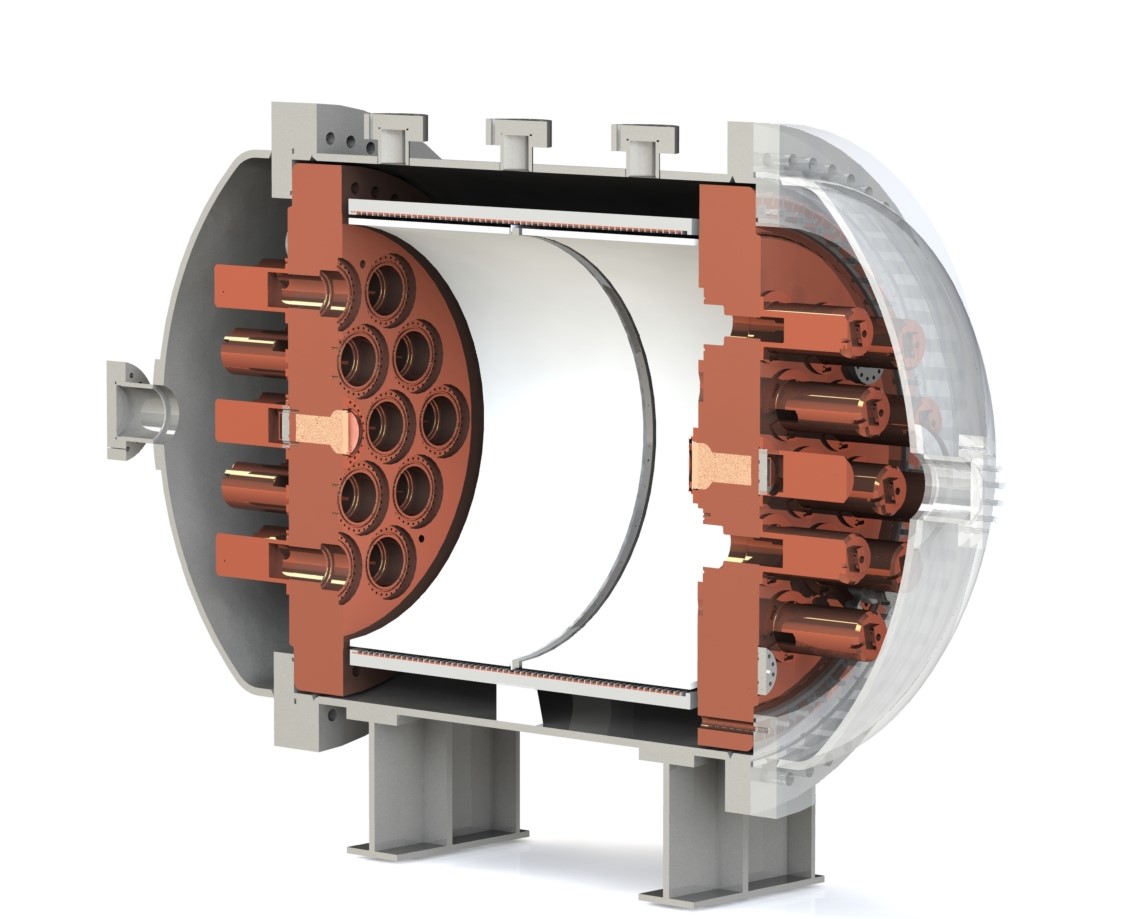}
    \includegraphics[width=0.4\textwidth]{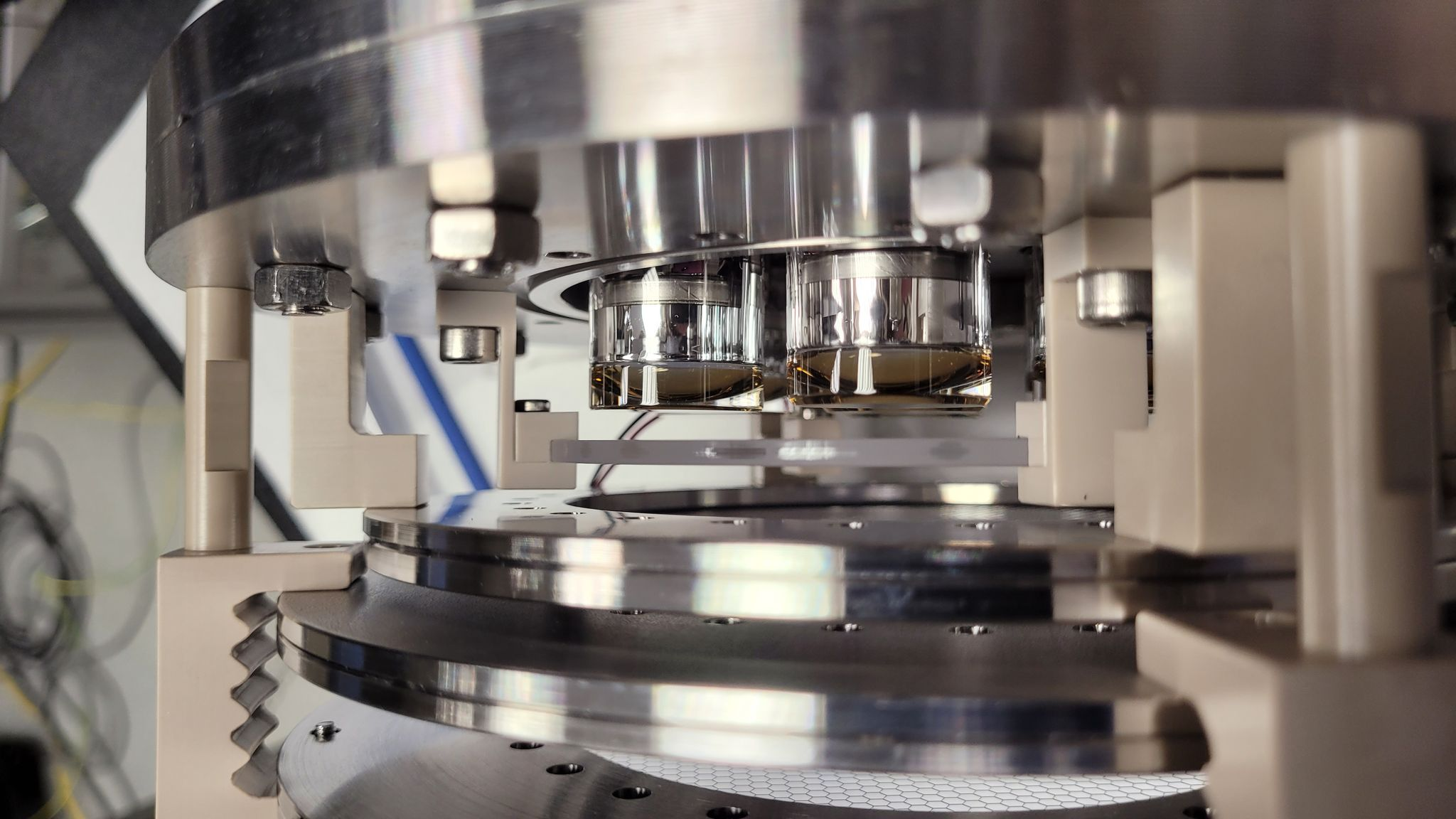}
    \caption{Left: Design of the GanESS experimentL: a symmetric TPC with the cathode at the center. Light is detected with PMTs at each side of the chamber. Right: A close-up of GaP's time projection chamber.}
    \label{fig:ganess}
\end{figure}

The GanESS experiment, illustrated in Fig.~\ref{fig:ganess}, will use of a high-pressure noble gas time projection chamber to measure \cenuns\ at ESS. The technique, which was developed for neutrinoless double beta searches, offers several interesting aspects for the detection of the process. First, it allows to operate with different nuclei with minimal modifications of the system. Concretely it will operate with Ar, Kr and Xe, providing the first ever \cenuns\ observations in Kr and Xe. Operation with different gases allows to explore different regions of the Non-Standar Interactions (NSI) of the neutrino-quarks phase space. A combined analysis of the data from different nuclei with same systematics can further reduce such allowed region improving total sensitivity to NSI \cite{Baxter:2019mcx,ABELE20231}. Second, amplification with electroluminiscence is widely used in particle physics and can easily produce several hundreds of photons per electron. A detector with a moderate optical coverage can thus be sensitive to a single free electron crossing the amplification region allowing to detect tiny energy depositions in the detector, as low as $\sim$25 eVee. 

Gaseous detectors weren't considered for \cenuns\ detection because their relatively low density (in comparison with solid scintillators or liquid phase detectors) could limit the statistics in a moderate size detector. Fortunately, the strong neutrino flux at the ESS solves this issue allowing full exploitation of the advantages of the noble gas TPC technology. Importantly, gaseous detectors avoids the liquid-gas interface  of dual phase detectors, where trapped electrons temporarily accumulate and end up producing a background signal increasing the detection threshold.

This technique is extraordinarily promising for \cenuns\  detection  albeit characterization of the response to few-keV nuclear recoils will be necessary. With this goal, we are currently comissioning GaP, a small prototype capable of operating up to 50 bar. GaP will serve to fully evaluate the low energy response of the technique, with a strong focus on measuring the quenching factor for the different noble gases that will later be used at GanESS. The GaP detector, shown on Fig.~\ref{fig:ganess}, is a 2 cm drift TPC currently equipped with 7 photomultiplier tubes near the amplification region. It recently started operations operating with 9.5 bar of Ar. The drift region is  being enlarged to 10 cm to maximize the sensitive volume. Once characterization of Ar is complete, Xe and Kr will follow.

\section{C$^{\circ}\!$sI: cryogenic pure CsI.}

The C$^{\circ}\!$sI detector is a natural evolution from the first \cenuns\ measurement (CsI[Na] at SNS \cite{COHERENT:2017ipa}). Instead of sodium-doped crystals, it will use pure CsI crystals cryogenically operated. Concretely, 7 crystals of 7$\times$7$\times$32 cm will be employed, adding up to $\sim$ 50 kg of mass.

\begin{figure}[!h]
    \centering
    \includegraphics[width=0.45\textwidth]{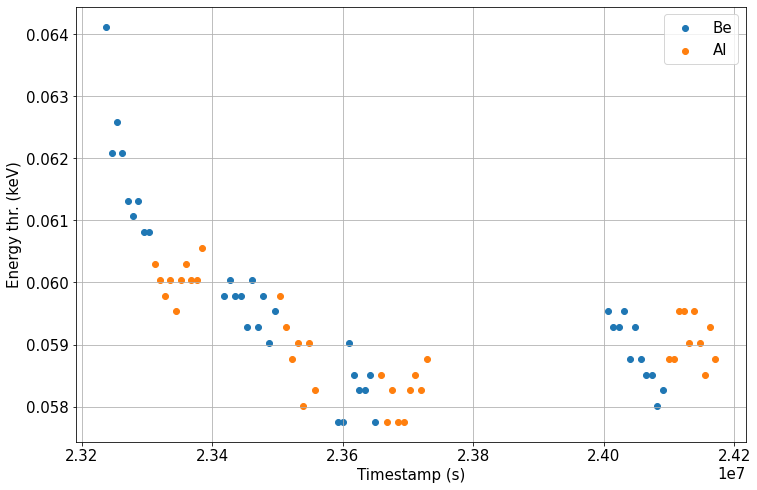}
    \includegraphics[width=0.45\textwidth]{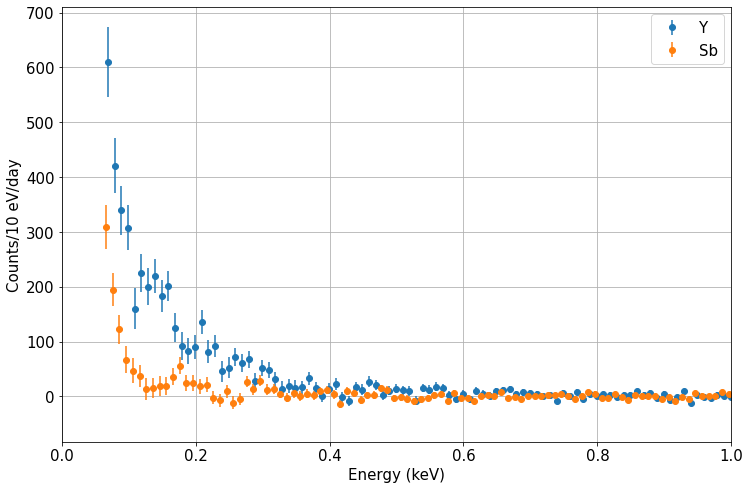}
    \caption{Left: Energy threshold of a few c.c.\ undoped CsI crystal operated at liquid nitrogen temperature. Light is detected with a LAAAPD following wavelength shifting with NOL-9. Right: energy spectrum from nuclear recoils induced in such crystal by $^{88}$Y/Be and $^{124}$Sb/Be photoneutron sources. Recoils from $^{124}$Sb/Be have a maximum recoil energy of $\sim$0.7 keVnr.}
    \label{fig:cosi}
\end{figure}

The reason for using undoped CsI comes from the fact that the light yield of such crystals increases enormously at cold temperatures (see Fig.~\ref{fig:cosi}), providing a two-fold increase on the light production when compared to CsI[Na] crystals. This light increase is fully exploited by wavelength shifting the emission of cryogenic CsI ($\sim$350 nm) to $\sim$600 nm with nanostructured organosilicon luminophores (NOL) and having the light read with a large area avalanche photodiode (LAAPD), with a quantum efficiency of $\sim$80\% at such wavelength. 

The combination of light yield, wavelength shifter and high quantum efficiency pushes down the detection threshold well below the keVee. In fact, a threshold slightly above 60 eV has been already achieved in a small (few c.c.) demonstrator at University of Chicago, with sub-keV nuclear recoils observed for the first time in a CsI crystal (Fig.~\ref{fig:cosi}). A dependance of the detection threshold with the operation temperature was observed, caused by variations in the LAAPD gain. Good temperature stability will be required to maximize the performance of the upcoming C$^{\circ}\!$sI detector.

\section{P-type HPGe detectors.}

Germanium detectors are being broadly used to detect \cenuns\ from reactor neutrinos and they are just as interesting for \cenuns\ from spallation sources. At least two p-type point contact high purity germanium detectors, each with $\sim$ 3 kg mass, will be deployed in the ESS for that purpose. One was used in the first observation of \cenuns\ in reactor neutrinos \cite{Colaresi:2022obx} and demonstrated a threshold of $\sim$180 eVee. A series of improvements have been identified and are currently under development to push the threshold below 100 eVee. A new detector is currently under construction, building on the experience of the Dresden detector, aiming at a threshold as low as 80 eVee. 

Existing discrepancies in the quenching factor of Ge need to be addressed to fully understand the \cenuns\ signal. These will be clarified in the coming months using thermal neutron capture. The gamma-induced recoils from this process will allow to measure the germanium quenching factor in the few hundred eV range.

\end{document}